\documentclass[useAMS,usenatbib]{mn2e}
\usepackage{graphics}
\usepackage{epsf}
\usepackage{subfigure}

\begin{document}

\title[Cepheid PC and AC relations]{Period-color and amplitude-color relations in classical Cepheid variables}
\author[Kanbur \& Ngeow]{Shashi M. Kanbur$^{1}$\thanks{E-mail: shashi@fcrao1.astro.umass.edu} and Chow-Choong Ngeow$^{1}$ 
\\
$^{1}$Department of Astronomy, University of Massachusetts      
\\
Amherst, MA 01003, USA}


\date{This is a preprint of an Article accepted for publication in MNRAS\copyright 2004 The Royal Astronomical Society}

\maketitle

\begin{abstract}
In this paper we analyze the behavior of Galactic, LMC and SMC Cepheids in
terms of period-color (PC) and amplitude-color (AC) diagrams at the phases of maximum, mean and minimum light. 
We find very different behavior between Galactic and Magellanic Cloud Cepheids. Motivated by the recent report of a break in LMC PC relations at 10 days (Tammann et al. 2002), we use the F-statistical test to examine the PC relations at mean light in these three galaxies. The results of the F-test support the existence of the a break in the LMC PC(mean) relation, but not in the Galactic or SMC PC(mean) relations. Furthermore, the LMC Cepheids also show a break at minimum light, which is not seen in the Galactic and SMC Cepheids. We further discuss the effect on the period-luminosity relations in the LMC due to the break in the PC(mean) relation.

\end{abstract}



\section{Introduction}

\citet[][hereafter SKM]{sim93} used hydro-dynamical
models to explain the observations of \citet{cod47}: Galactic
Cepheids show a spectral type independent of period at maximum light
and a spectral type at minimum light that gets later as the period increases.
SKM computed radiative hydro-dynamical models of Galactic Cepheids which agreed
with Code's observations. SKM interpreted these observational phenomena as being due to the
location of the photosphere relative to the hydrogen ionization front. They
further used the Stefan-Boltzmann law applied at maximum and minimum light to show that

\begin{eqnarray}
\log T_{max} - \log T_{min} = {1\over{10}}(V_{min} - V_{max}),
\end{eqnarray}

where $T_{max}$ and $T_{min}$ refer to the effective temperature at maximum and minimum light, respectively. 
Thus, higher optical amplitudes are associated with higher temperature amplitudes, which are in turn
related to higher values of $T_{max}$ and/or $T_{min}$. For this study we do not assert a
causal relation between higher temperatures and higher amplitudes preferring to
refer to these quantities as being ``associated''.
If, for some reason, either $T_{max}$ or
$T_{min}$ does not increase as the amplitude increases, equation (1)
predicts a relationship between amplitude and $T_{min}$ or $T_{max}$ respectively.
SKM used data from \citet{pel76} and \citet{mof80,mof84} to show that Galactic Cepheids
are such that higher amplitude stars are driven to
cooler temperatures at minimum light. This, according to equation (1), is because the range of temperatures at maximum
light is independent of period for a large range of periods. So the form of the
period-color (PC) relation at maximum light is related to the form of the amplitude-color (AC) relation
at minimum light, and vice versa.

\citet{tam03} used Galactic and OGLE LMC/SMC Cepheid
data to show that there is a difference in the PC relations in these
three galaxies. Furthermore, \citet{tam02a} and \citet{tam02} show the existence of two PC
relations, one for short ($P<10$ days) and one for long ($P>10$ days) period
Cepheids, in the LMC. Motivated by this and the presence of high quality Magellanic Cloud Cepheid data from the
OGLE project \citep{uda99a}, we decided to investigate the properties of 
Magellanic Cloud Cepheids in terms of their PC and AC relations at maximum, mean and minimum light. 

In addition to the two major reasons mentioned above, we list a number of other arguments motivating the present study:

\begin{enumerate}
\item   Since mean light is just that - the average over a range of
values - interesting properties of Cepheids at mean light are due to the average of
these properties at all pulsation phases. By investigating the phases of maximum and minimum light, we
are studying those phases of
stellar pulsation which contribute to the observed
properties at mean light.
Our interest lies in understanding breaks in the LMC mean light Cepheid PC and period-luminosity (PL) relations
at 10 days reported by \citet{tam02a}.
Let $y_{ij}$ be the (absolute) magnitude of
Cepheid variable stars, $i=1,...,n_{star}$ at the $j^{th}$ phase ($j=1,...,N$) during a pulsation period $P_i$. Then we can formulate a PL relation at a particular phase as,

\begin{eqnarray}
y_{ij} = a_j + b_j \log (P_i),
\end{eqnarray}

where $a_j,b_j$ are the unknown coefficients as a function of the phase $j$.
If we define $y_i$ as ${\sum_{j=1}^{j=N}y_{ij}}/N$, the average over the pulsation period, it is easy to show that
$y_i = a_m + b_m \log (P_i),$
where $a_m,b_m$ are the average over phase of the $a_j,b_j$ in equation (2)\footnote{Note that $b_m$ may not lie in between $b_{max}$ and $b_{min}$, the slopes at maximum and minimum light, respectively. A similar conclusion also holds for the zero-point, $a_m$.}.
This will be true if $y$ is measured in intensities and then the intensity
mean converted to magnitudes or if $y$ is measured in magnitudes. Of course
the magnitude mean and intensity means are in general not equal to each
other but the difference is small ($\sim0.03mag.$) and constant
over a wide range of periods (for example, see \citealt{gie98}). Consequently, some insight into the mean light relation can be
gained by studying PL relations at various phases, e.g. at maximum and minimum light. Since the PC relation affects the PL relation (see, e.g.,  \citealt{mad91} for the basic physics of PC and PL relations), 
it is of interest to study the PC relation at various phases.
Furthermore, the maximum and minimum light are closely associated with the more interesting phases of stellar pulsation: the expansion/contraction velocity is close to its maximum value when the photosphere is passing through the mean radius of the star (see, e.g., \citealt{mih03} for the details).

\item   The amplitude is a fundamental observational and theoretical quantity in stellar
pulsation. Kanbur and Ngeow (2004, in-preparation) show that the
amplitude is a very good descriptor of light curve shape, and the V band amplitudes are correlated to the first Principal Component ($PC1$) of the light curve. In addition, \citet{kan02} demonstrated that $PC1$ can explain over 90$\%$ of the
variation in light curve shape. Thus the V band amplitude can be taken to be a good descriptor of V band light curve
shape. A similar conclusion holds for the I band. 
Because the optical brightness fluctuations of Cepheids are predominantly due to temperature fluctuations \citep{cox80}, it is thus
instructive to examine AC diagrams at maximum and minimum V band light. Furthermore, since AC relations are related to PC
relations through equation (1), their study can serve as a useful
complement to strengthen any conclusions reached using PC relations. 

\end{enumerate}


\section{The Data}

        The standard Johnson-Cousins V and I band data for the fundamental mode Cepheids in the Galaxy, LMC and SMC were used in this study. For both of the LMC and SMC Cepheids, the periods, V and I band photometric data and the $E(B-V)$ values for every Cepheid were taken from the OGLE \citep{uda99b,uda99c} website\footnote{\texttt{http://bulge.princeton.edu/$\sim$ogle/}}. There are 771 and 488 fundamental mode Cepheids (as classified by OGLE team) in LMC and SMC, respectively. For Galactic Cepheid data, we considered only the Cepheids classified as ``DCEP'' in the General Catalog of Variable Stars (GCVS, fourth edition, \citealt{kho98}). The periods for the Galactic Cepheids are taken from the McMaster Cepheid database\footnote{\texttt{http://dogwood.physics.mcmaster.ca/Cepheid//HomePage.html}}, and the $E(B-V)$ values are adopted from \citet[][listed as $E(B-V)_{corr}$ in their table 1]{tam03}. The photometric data for Galactic Cepheids were obtained from two sources: (a) \citet{mof84}, where actual data were downloaded from the McMaster Cepheid database; and (b) \citet{ber97}\footnote{Note that the column of $R-I_c$ should be swapped with column of $V-I_c$ in this dataset.} and \citet{ber01}. For Cepheids that have entries in both \citet{ber97} and \citet{ber01}, photometric data from these two sources were merged together to provide a better light curve. Since the bandpasses used in \citet{mof84} are in Johnson V and I, we converted the Johnson I band photometric data in this dataset to Cousins I band with the color transformations given in \citet{cou85}.

        The Cepheid photometric data in these three galaxies were then fitted with a Fourier expansion of the following form \citep{sch71,nge03}:

        \begin{eqnarray}
        m(t) & = & m_0 + \sum^{M}_{i=1} [A_i\cos(2\pi i\Phi(t) + \phi_i)], 
        \end{eqnarray}  

        where $\Phi(t)=(t-t_0)/P - int[ (t-t_0)/P ]$, with $t_0$ being a common starting epoch for all Cepheids in both bands. We used a simulated annealing technique to fit the data with this Fourier expansion, as described in \citet{nge03}. Thus our Fourier fits are carried out to published photometric data and our means are magnitude means. For Galactic data, we adopted a fifth order expansion ($M=5$) to most of the Cepheids. However, in certain cases a fourth or sixth order Fourier expansion gave a better fit to the data. For OGLE LMC Cepheids, we fit the data with $M=4$, as this dataset is also used in \citet{nge03} and in \citet{kan03}. For OGLE SMC Cepheids, we mainly fit fourth or fifth order Fourier expansions to the data, while some of them were fitted with sixth order. All fitted light curves were then visually inspected (see Kanbur et al. 2003 for the selection criteria). Figure 7 of \citet{nge03} illustrates the improvement that can be obtained when using our fitting method to the observed light curves. We only selected those Cepheids with well fitted light curves in {\it both} V and I bands. In addition, we exclude Cepheids with $\log (P)<0.4$ to avoid possible contamination from first overtone Cepheids \citep{uda99a}. The numbers of Cepheids in the final samples are: 79 from \citet{mof84} data; 75 from Berdnikov data; 634 from OGLE LMC data; and 391 from OGLE SMC data. From the Fourier fits we obtained the following quantities:

\begin{enumerate}
\item   V and I band amplitude from the numerical maximum and minimum of the Fourier fits:
$V_{amp}=V_{min}-V_{max}$, $I_{amp}=I_{min}-I_{max}$.
\item   $(V-I)_{max}$: defined as $V_{max} - I_{phmax}$, where $I_{phmax}$ is the I band 
magnitude at the same phase as $V_{max}$.
\item   $(V-I)_{mean}$: defined as $V_{mean} - I_{phmean}$, where $I_{phmean}$ is the I band
magnitude at the same phase as $V_{mean}$. $V_{mean}$ is the V band magnitude closest to $m_0$, the
mean value in equation (3).
\item   $(V-I)_{min}$: defined as $V_{min} - I_{phmin}$, where $I_{phmin}$ is the I band
magnitude at the same phase as $V_{min}$.
\end{enumerate}

        Finally, the colors at these three phases have been corrected for extinction using $A_{V,I}=R_{V,I}E(B-V)$. The values of $R$ are: $R_V=3.17$, $R_I=1.89$ for Galactic data \citep{tam03}, and $R_V=3.24$, $R_I=1.96$ for OGLE LMC and SMC data \cite{uda99b,uda99c}. In fact, the results are unchanged for the values of $R_V$ and $R_I$ as long as $\Delta R\equiv R_V-R_I=1.28$ as given in \citet{tam03}, because $(V-I)_o=(V-I)-\Delta R\ E(B-V)$. We apply the same extinction values of $A_{V,I}$ to the colors at maximum, mean and minimum light, since the quantities of $A_{V,I}$ should remain unchanged for any pulsation phases. In addition, our results are unchanged if we adopt $(<V>-<I>)_o$ for the mean color as in item (iii) above.

 

        \begin{figure*}
        \vspace{0cm}
        \hbox{\hspace{0.2cm}\epsfxsize=8.5cm \epsfbox{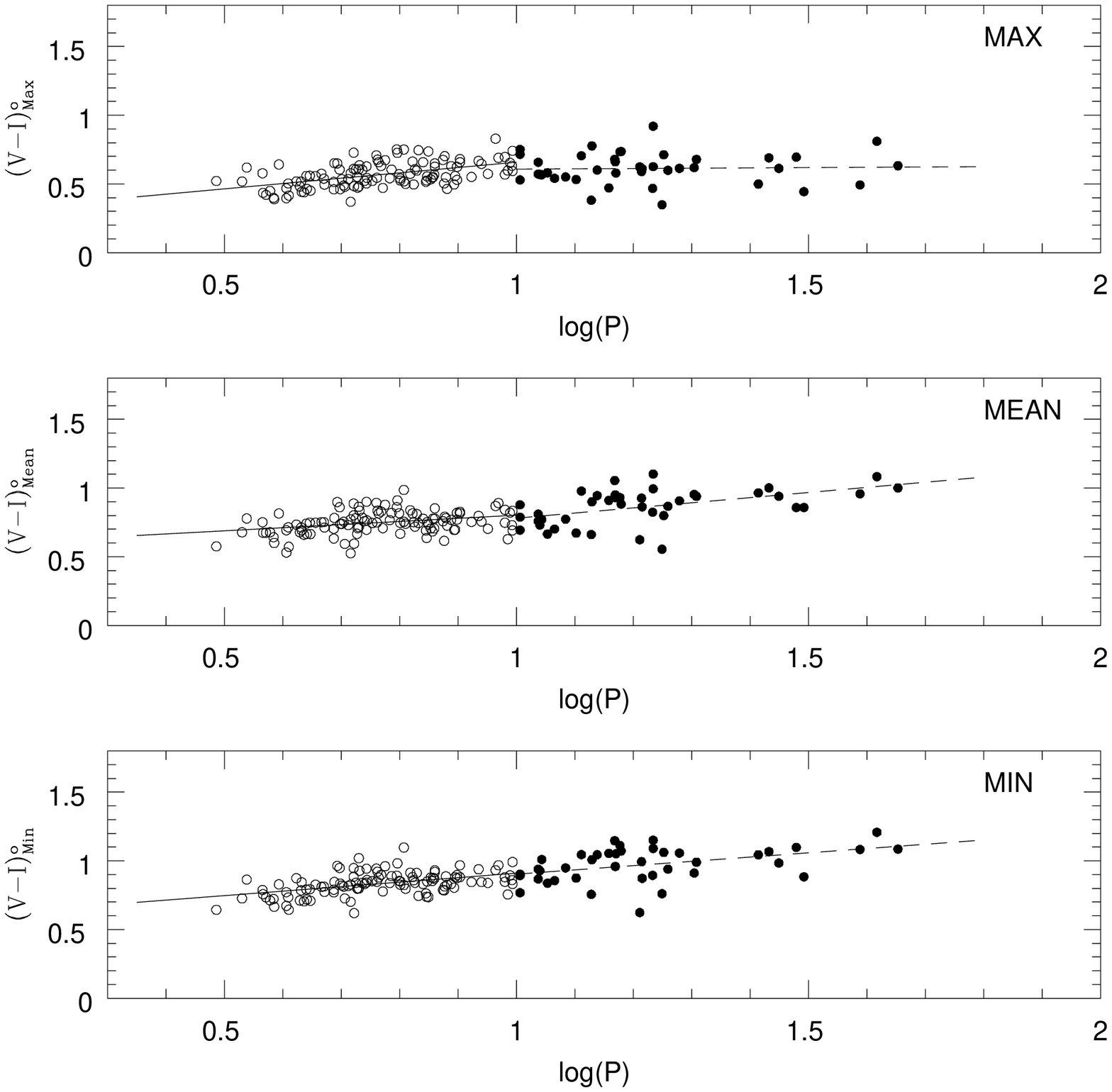}
        \epsfxsize=8.5cm \epsfbox{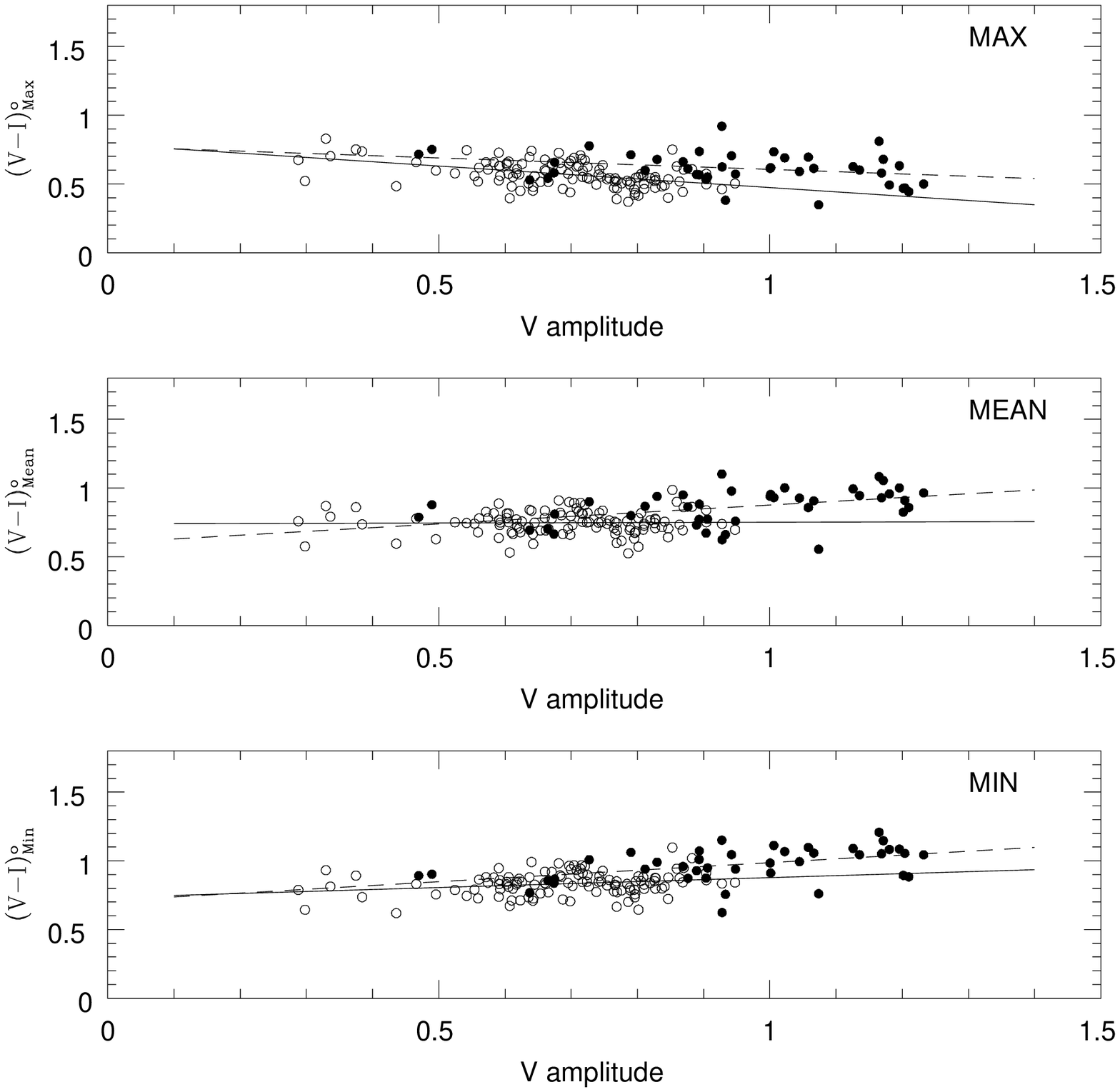}}
        \vspace{0cm}
        \caption{The period-color and amplitude-color relations at phases of maximum, mean and minimum light for Galactic Cepheids. The open circles and filled circles correspond to short and long period Cepheids, respectively. The dashed and solid lines are the fitted period-color and amplitude-color relations to short and long period Cepheids, respectively. The number of short and long period Cepheids are $n_{short}=113$ and $n_{long}=41$. {\it Left} (a): The period-color relations. {\it Right} (b): The amplitude-color relations. \label{gal}}
        \end{figure*}


        \begin{figure*}
        \vspace{0cm}
        \hbox{\hspace{0.2cm}\epsfxsize=8.5cm \epsfbox{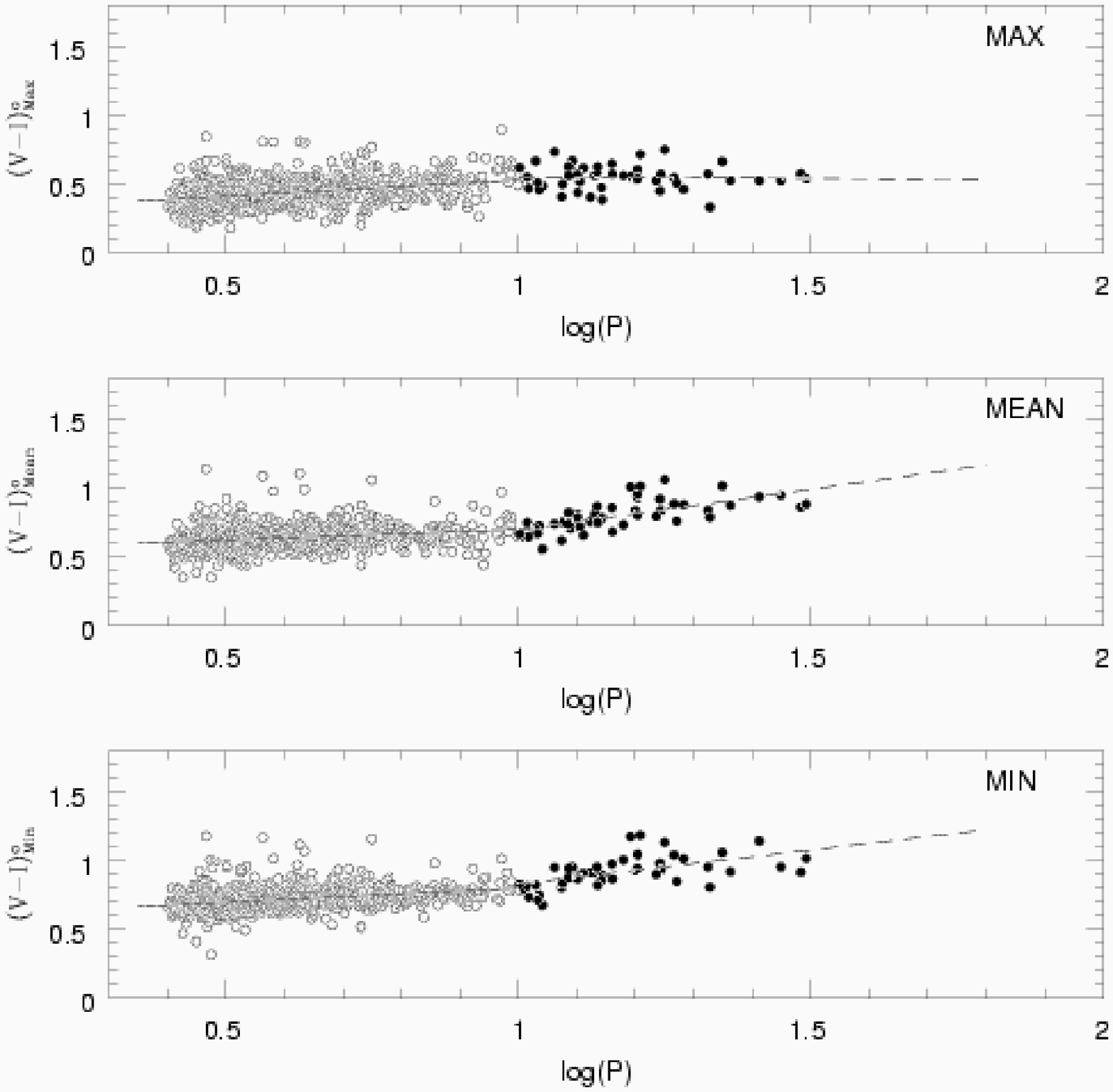}
        \epsfxsize=8.5cm \epsfbox{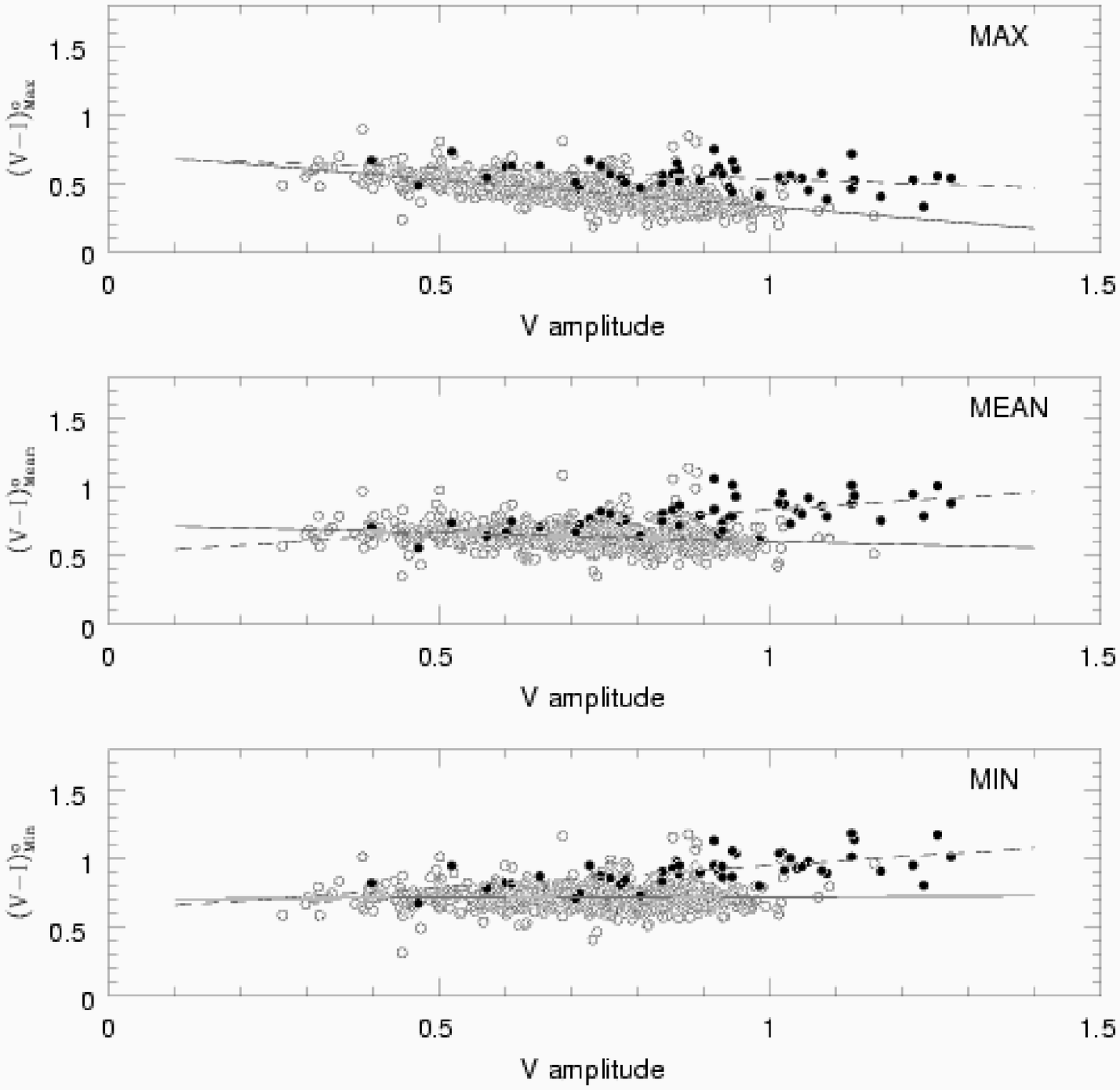}}
        \vspace{0cm}
        \caption{The period-color and amplitude-color relations at phases of maximum, mean and minimum light for OGLE LMC Cepheids. The open circles and filled circles correspond to short and long period Cepheids, respectively. The dashed and solid lines are the fitted period-color and amplitude-color relations to short and long period Cepheids, respectively. The number of short and long period Cepheids are $n_{short}=585$ and $n_{long}=49$. {\it Left} (a): The period-color relations. {\it Right} (b): The amplitude-color relations. \label{lmc}}
        \end{figure*}


        \begin{figure*}
        \vspace{0cm}
        \hbox{\hspace{0.2cm}\epsfxsize=8.5cm \epsfbox{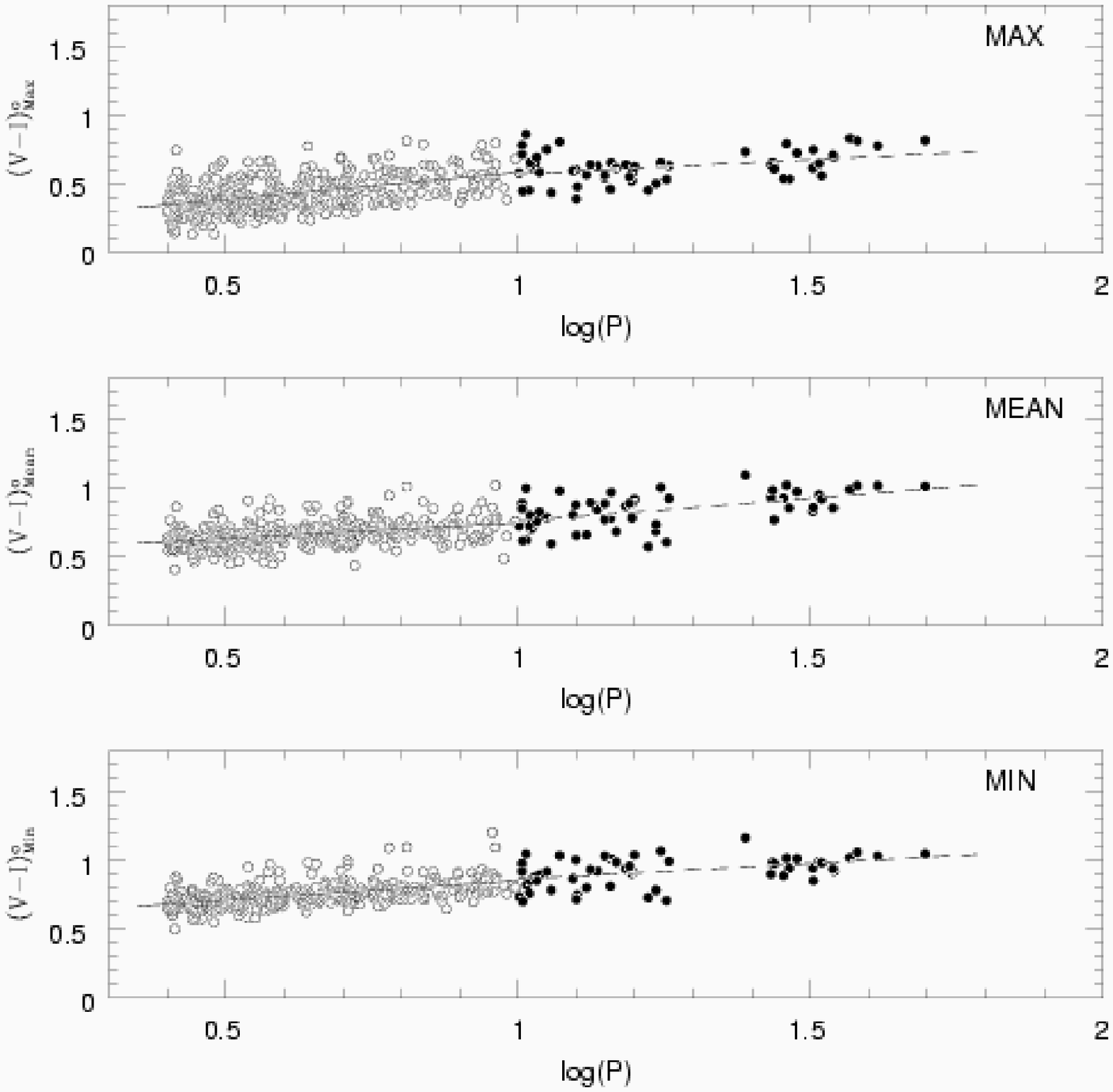}
        \epsfxsize=8.5cm \epsfbox{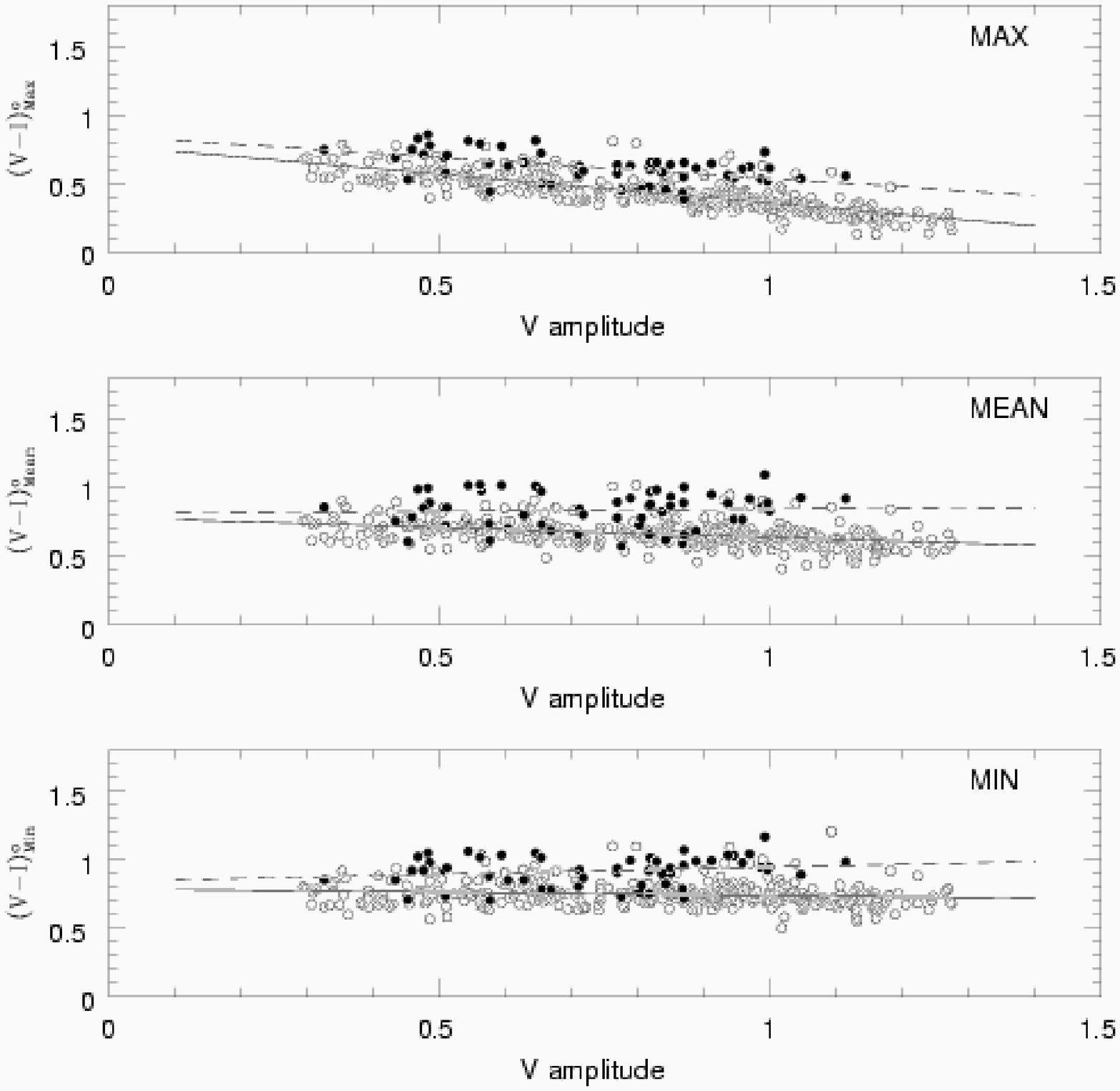}}
        \vspace{0cm}
        \caption{The period-color and amplitude-color relations at phases of maximum, mean and minimum light for OGLE SMC Cepheids. The open circles and filled circles correspond to short and long period Cepheids, respectively. The dashed and solid lines are the fitted period-color and amplitude-color relations to short and long period Cepheids, respectively. The number of short and long period Cepheids are $n_{short}=334$ and $n_{long}=57$. {\it Left} (a): The period-color relations. {\it Right} (b): The amplitude-color relations. \label{smc}}
        \end{figure*}

\section{Analysis and Results}

Figures \ref{gal}, \ref{lmc} and \ref{smc} show the plots of $\log (P)$ vs. $(V-I)_o$ color and $V_{amp}$ vs. $(V-I)_o$ color at maximum, mean and minimum light (see the definition in Section 2) for the Cepheids in Galaxy, LMC and SMC respectively.
Open and closed circles in these figures are for short ($\log (P) < 1.0$) and long ($\log (P) > 1.0$) period Cepheids, respectively.
In what follows short and long period Cepheids are always used in this sense.
For all three data sets in the Galaxy, LMC and SMC, we can fit linear relations
between $\log (P)$ and $(V-I)_o$ color and $V_{amp}$ and $(V-I)_o$
color, both for the entire sample, and also for short and long period Cepheids separately. 
Tables \ref{tab1} and \ref{tab2} show these results for the fits to period-color and amplitude-color relations, respectively. 
In these tables, column 1 gives the phase at which the fit is made - either at
maximum, mean or minimum light. Column 2 is labelled $MRSS$ and represents the mean residual sum of
squares from the fit to the entire sample. The slope of the linear fit together with its associated error is given
in column 3. Columns 4 and 5 give the residual mean sum of squares and slope plus error for the
short period Cepheids. Columns 6 and 7 do this for the longer period Cepheids. We discuss columns 
8-11 shortly. In Figures \ref{gal}-\ref{smc} we have plotted the fitted lines for short and long period
Cepheids as solid and dashed lines respectively.

What is of interest for the present study is whether these PC plots show statistical evidence 
of a change of slope at a period of 10 days. To investigate this, we can fit
a regression line over the entire period range and then fit two regression lines,
one separately for short and long period Cepheids. The former case is the reduced model
while the latter case with two separate regression lines is the full model.
We can write the model under consideration as

\begin{eqnarray}
Y = a_S W_S + a_L W_L + b_S Z_S + b_L Z_L + \epsilon.
\end{eqnarray}

Here $Y$ is the dependent variable, in our case $(V-I)_o$ color at any of the three
phases under consideration. $W_S$ is an indicator variable which is 1 if the star's period
is less than 10 days and 0 otherwise. $W_L$ is an indicator variable which is 0 if the
star's period is less than 10 days and 1 otherwise. The variable $Z_S$ contains the
independent variable, either $\log (P)$ or $V_{amp}$, but is zero if the
period is greater than 10 days. $Z_L$ is similar but is zero for periods
less than 10 days. The parameters $a_S$, $a_L$, $b_S$ and $b_L$ are the zero-points and slopes for short and long period Cepheids, respectively. Thus what is of interest is if the data are consistent with
$b_L = b_S$: the slope is same for long and short
period Cepheids. \citet{wei80} shows that in this situation an appropriate
F test statistic can be formulated as described in the following equation,

\begin{eqnarray}
F = \frac{(RSS_R - RSS_F)/[(n-2)-(n-4)]}{RSS_F/(n-4)},
\end{eqnarray}

where $RSS_R, RSS_F$ are the residual sums of squares in the reduced (single line regression) and full
model (two lines regression), respectively, and $n$ is the number of stars in the entire sample. We refer the F
statistic in equation (5) to
an F distribution, $F_{2,n-4}$, under the null hypothesis that the two-parameter 
regression (i.e. a single line) is sufficient. The four-parameter
regression (i.e. two lines) will have a smaller residual sum of squares. The probability
of the observed value of the F statistic, $P(F)$, under the null hypothesis, gives the significance of this reduction
in sum of squares. Thus, a ``large'' value of F indicates that the null hypothesis can be rejected. Column 8 of Tables \ref{tab1} and \ref{tab2} gives the values of the F statistic and column 9 gives its probability value
as described above. Note that because there are fewer observed long period Cepheids, the error on the slope
for the long period PC relations is generally larger but this is automatically taken account of by the F test.

It could also be that the short and long period slopes are similar but that the F statistic produces
a significant result because the intercepts are different. In this case,
we can also
compare the statistical significance of differences in the slopes of these regressions by referring the
quantity,

\begin{eqnarray}
t &=& (b_S - b_L) \frac{1}{S_{b_S-b_L}},  \\
S_{b_S-b_L} &=& \sqrt{\frac{MS}{SS_S} + \frac{MS}{SS_L}}, \\
MS &=& \frac{SS_S + SS_L}{n_S + n_L - 4}
\end{eqnarray}

to a $t$ distribution with $n_S + n_L - 4$ degrees of freedom. $P(t)$ is the
probability of the observed value of the $t$ statistic under the null hypothesis
that the two slopes are equal. In the above formulae
$SS_S, SS_L, n_S, n_L, b_S, b_L$ are the residual sum of squares, the number of
Cepheids and the slopes in the PC or AC relation for short and
long period Cepheids, respectively. Columns 10 and 11 in Tables \ref{tab1} and \ref{tab2} give the values of $t$ statistic and 
its associated $P(t)$ value from two-tail $t$ distribution.


        \begin{table*}
        \centering
        \caption{Period-color relations at maximum, mean and minimum light.}
        \label{tab1}
        \begin{tabular}{lcccccccccc} \hline
        Phase & $MRSS_{all}^a$  & $Slope_{all}^b$  & $MRSS_S^a$ & $Slope_S^b$ & $MRSS_L^a$ & $Slope_L^b$ & F & P(F) & t & P(t) \\
        (1)   & (2)   & (3)            & (4)   & (5)    & (6)   & (7)    & (8) & (9) & (10) & (11) \\ \hline \hline
              &       &                &       &        & Galactic &     &   &      &   &      \\ \hline
        Max   &0.009 &$0.142\pm0.031$ &0.006 &$0.389\pm0.062$ &0.013 &$0.020\pm0.105$  &7.95 &0.00 &3.40 & 0.00 \\
        Mean  &0.008 &$0.264\pm0.030$ &0.006 &$0.229\pm0.062$ &0.013 &$0.373\pm0.104$  &1.05 &0.35 &1.34 & 0.18 \\
        Min   &0.007 &$0.310\pm0.028$ &0.005 &$0.330\pm0.058$ &0.013 &$0.309\pm0.104$  &0.06 &0.94 &0.21 & 0.83 \\
        \hline
              &       &       &       &        & OGLE LMC &     &   &      &   &      \\ \hline
        Max   &0.010 &$0.205\pm0.020$ &0.010 &$0.234\pm0.030$ &0.008 &$-0.031\pm0.101$ &2.67  &0.07 &2.24 & 0.03 \\
        Mean  &0.009 &$0.239\pm0.018$ &0.008 &$0.152\pm0.027$ &0.007 &$0.590\pm0.097$  &13.19 &0.00 &4.10 & 0.00 \\
        Min   &0.008 &$0.291\pm0.018$ &0.008 &$0.204\pm0.027$ &0.009 &$0.493\pm0.109$  &10.38 &0.00 &2.73 & 0.01 \\
        \hline
              &      &       &       &        & OGLE SMC &     &   &      &   &      \\ \hline
        Max   &0.013 &$0.324\pm0.021$ &0.013 &$0.396\pm0.039$ &0.011 &$0.207\pm0.071$ &3.16  &0.04 &2.23 & 0.03 \\
        Mean  &0.008 &$0.272\pm0.017$ &0.007 &$0.229\pm0.030$ &0.012 &$0.340\pm0.074$ &1.77  &0.17 &1.65 & 0.10 \\
        Min   &0.007 &$0.276\pm0.015$ &0.006 &$0.282\pm0.027$ &0.010 &$0.229\pm0.066$ &0.42  &0.66 &0.87 & 0.38 \\
        \hline
        \end{tabular}
        \begin{list}{}{}
        \item   $^a$ $MRSS$ denotes mean of residual sum of squares from the fit. $MRSS_{all}$, $MRSS_S$ and $MRSS_L$ are for the fits to all, short and long Cepheids in the sample, respectively.
        \item   $^b$ Slopes for the period-color relations for all, short and long period Cepheids in the sample.
        \end{list}
        \end{table*}

        \begin{table*}
        \centering   
        \caption{Amplitude-color relations at maximum, mean and minimum light.}
        \label{tab2}
        \begin{tabular}{lcccccccccc} \\ \hline
        Phase & $MRSS_{all}^a$ & $Slope_{all}^b$ & $MRSS_S^a$ & $Slope_S^b$ & $MRSS_L^a$ & $Slope_L^b$ & F & P(F) & t & P(t) \\ 
        (1)   & (2)   & (3)            & (4)   & (5)    & (6)   & (7)    & (8) & (9) & (10) & (11) \\ \hline \hline
              &      &       &       &        & Galactic &     &   &      &   &      \\ \hline
        Max   &0.010 &$-0.098\pm0.042$ &0.007&$-0.313\pm0.058$ &0.012 &$-0.169\pm0.088$  &15.54 & 0.00 & 1.50 & 0.14 \\
        Mean  &0.010 &$0.241\pm0.042$  &0.007 &$0.011\pm0.059$  &0.014 &$0.275\pm0.095$  &10.98 & 0.00 & 2.65 & 0.01 \\
        Min   &0.009 &$0.322\pm0.041$  &0.007 &$0.145\pm0.058$  &0.013 &$0.276\pm0.089$  &9.76  & 0.00 & 1.36 & 0.18 \\
        \hline
              &      &       &       &        & OGLE LMC &     &   &      &   &      \\ \hline
        Max   &0.010 &$-0.285\pm0.023$ &0.007 &$-0.393\pm0.023$ &0.007 &$-0.160\pm0.058$ &93.12 & 0.00 & 3.62 & 0.00 \\
        Mean  &0.011 &$0.016\pm0.025$  &0.008 &$-0.118\pm0.024$ &0.008 &$0.322\pm0.064$  &95.99 & 0.00 & 6.42 & 0.00 \\
        Min   &0.011 &$0.140\pm0.025$  &0.009 &$0.017\pm0.025$  &0.009 &$0.321\pm0.065$  &92.66 & 0.00 & 4.33 & 0.00 \\
        \hline
              &      &       &       &        & OGLE SMC &     &   &      &   &      \\ \hline
        Max   &0.010 &$-0.438\pm0.021$ &0.007 &$-0.419\pm0.018$ &0.010 &$-0.308\pm0.070$ &80.45 & 0.00 & 1.79 & 0.07 \\
        Mean  &0.012 &$-0.158\pm0.023$ &0.007 &$-0.144\pm0.019$ &0.017 &$0.034\pm0.092$  &75.40 & 0.00 & 2.54 & 0.01 \\
        Min   &0.012 &$-0.065\pm0.023$ &0.008 &$-0.047\pm0.020$ &0.011 &$0.104\pm0.076$  &82.45 & 0.00 & 2.21 & 0.03 \\
        \hline
        \end{tabular}
        \begin{list}{}{}
        \item   $^a$ $MRSS$ denotes mean of residual sum of squares from the fit. $MRSS_{all}$, $MRSS_S$ and $MRSS_L$ are for the fits to all, short and long Cepheids in the sample, respectively.
        \item   $^b$ Slopes for the amplitude-color relations for all, short and long period Cepheids in the sample.
        \end{list}
        \end{table*}

Our figures do not contain error bars on the estimated values of the $(V-I)_o$ color. These error
bars are typically less than the size of the symbol. This includes errors
due to photometry and reddening. A typical V band photometric error quoted by the OGLE team is
$0.015 mags$ and perhaps as much as $0.05mags$ for reddening. If we add these in quadrature, a rough
estimate of
sigma for the photometry is $0.052mags$. If we assume a similar figure for the I band (with errors of $\sim0.001mags.$ to $0.005mags$) and again
add up the errors in quadrature, then a typical error for the $(V-I)_o$ color is $\sim0.07 mags$. Strictly
speaking this error should be included in a weighted least squares fit when constructing
our period-color or amplitude-color fits. The weights are the inverse of the error
assigned to each data point. However if the weights are constant for each star then they cancel
out in the calculation of the F statistic and in the derivation of the coefficients of the
linear regression. Hence the only way that inclusion of errors in our least squares fits can affect
the significance of our F statistic results is if the errors are systematically greater for
large numbers of either short or long period Cepheids. If this is the case, it will affect not only
our work but also published values of the LMC PL relations in the V and I band (e.g., in \citealt{uda99a}).

\begin{figure}
        \hbox{\hspace{0.1cm}\epsfxsize=7.5cm \epsfbox{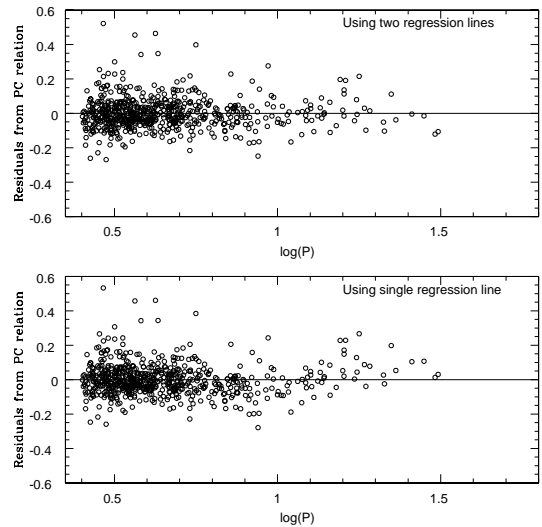}}
        \caption{Plots of the residuals of LMC period-color relation at mean light. {\it Top:} (a) The residuals from the long and short period PC relations as a function of $\log(P)$. {\it Bottom:} (b) The residuals from using a single PC relation as function of $\log(P)$. From the plots it is clear that the residuals are homoskedastic. In addition, the residuals from the fit with a single PC relation show a trend in long period Cepheids. This trend is reduced if two PC relations were used to fit the data.}
        \label{residual}
        \end{figure}

\begin{figure}
        \hbox{\hspace{0.1cm}\epsfxsize=7.5cm \epsfbox{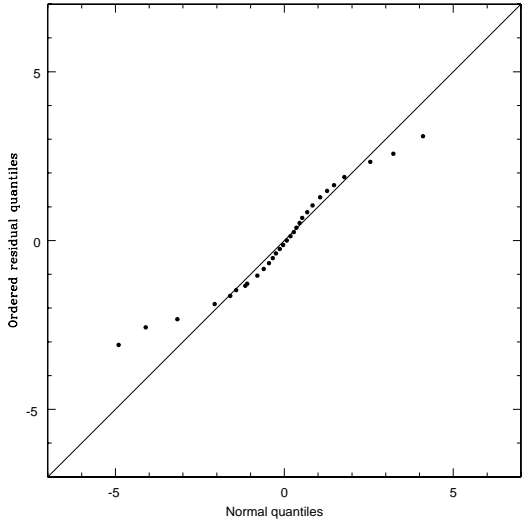}}
        \caption{Plot of the ordered residual quantiles vs. normal quantiles, the qq-plot. The line is the case of $y=x$.}
        \label{qq}
        \end{figure}

We now discuss the assumptions required by the F test. These are that 
the error ($\epsilon$) in equation (4) is constant for all the stars in our study, and further they are normally distributed with zero mean. To test these assumptions, we analyze the OGLE
LMC data in greater detail. Figure \ref{residual} shows the plots of residuals from a single PC
fit and two PC fits to the entire data against $\log(P)$. We see no discernible trend in the size of the residuals with period and
are led to the conclusion that the residuals are homoskedastic (i.e. $\sigma\sim$ constant). This supports
our approach of adopting a constant $\sigma$ to gauge photometric
errors when doing a least squares fit. In addition, the residuals in Figure \ref{residual} from the fit with a single PC relation show a trend (though not in the size of the residuals) in long period Cepheids, which is reduced when using two PC relations for short and long period Cepheids to fit the data. In order to test that the residuals are normally distributed, we can plot the quantiles of the distribution
resulting from the ordered residuals against the expected quantiles from a normal distribution:
a qq-plot. If the residuals are indeed normally
distributed then this plot should be close to the line
$y=x$. Figure \ref{qq} shows such a plot. There is some small departure from
normality at the extremes but we contend that this plot justifies our use of the F test.


There is a great deal of information in Table \ref{tab1} \& \ref{tab2}, but for the context of the present study 
we summarize the results in the following subsections.

\subsection{Results of period-color relations}

For the Galactic Cepheids, the slope of the long period PC relation at maximum light
is close to zero. Further, the overall slope at maximum light is the
shallowest. This then supports the work of \citet{cod47}. In addition, the  F and $t$ test indicates that the 
PC relation at maximum light is not consistent with a single slope.
However, the mean and minimum light PC relations for the Galactic Cepheids are consistent with a single line.

For the LMC Cepheids, the slopes of the long period PC relation at maximum light is also close to zero, decreasing significantly from its value for short
period Cepheids. The slope at mean and minimum light all increase when going from short to
long period Cepheids.  
In the LMC, we see that the PC relation at maximum light is not consistent with a single line at the 93$\%$ confidence
level.
However, in terms of the F test, the PC 
relation at mean and minimum light does show evidence to support a break at 10
days. Note that at both these phases, the slope of the
PC relation for longer period Cepheids is much steeper than for
shorter period Cepheids.

In the case of the SMC again the PC relation at maximum light is much flatter for longer period
Cepheids than for shorter period. The relation at mean light for longer periods is steeper than at
shorter period. Only the PC relation at maximum light shows statistical evidence of a break at
10 days.

We note that the slope of PC relation at maximum light always decreases in going from
short to long period Cepheids in all three galaxies.
In contrast, the slope at mean and minimum light always increases except for the Galactic and SMC Cepheids at minimum light. We also make the 
interesting observation that the dispersion of the PC relation, whether for the
entire sample or either short or long period group, is always the smallest at
minimum light in these three galaxies. A glance at the
F statistic for mean light for the LMC Cepheids shows that we can reject the null hypothesis
of a single PC relation for long and short period LMC Cepheids at greater than the 99$\%$ confidence level, confirming the finding of the broken PC relation in LMC \citep{tam02,tam02a}. However, the
Galactic and SMC mean light PC relation is consistent with one line.

\subsection{Results of amplitude-color relations}

The AC relations all show statistical evidence of two lines at greater than $99.9\%$ confidence level with the F test.
At mean and minimum light, we see that the slopes of the AC relation increase significantly when going from
short to long period Cepheids in all three galaxies. However at maximum light the slope becomes shallower for long period Cepheids. Also, the slopes of the AC relation at maximum light are always negative for both short and long periods Cepheids in Galaxy, LMC and SMC, which we do not see in the mean and minimum light. In addition, the slopes are flat for the short period Galactic Cepheids and long period SMC Cepheids at mean light, but not for the Cepheids in LMC. 

We also remark on some important differences in the AC plane between Galactic and Magellanic Cloud 
Cepheids. At minimum light in the LMC and SMC, the slope of the AC relation is $0.017\pm0.025$ and
$-0.047\pm0.020$, respectively, for
short period Cepheids, whereas it is $0.145\pm0.058$ for Galactic Cepheids - more than $2\sigma$
difference. For longer period Cepheids in the LMC and SMC, this slope becomes positive ($0.321\pm0.065$
and $0.104\pm0.076$), 
whilst the longer period AC slope in
the Galaxy is always significantly above zero.

Even though Table \ref{tab2} shows that AC relations are significantly different
for short and long period Cepheids in the SMC, Figure \ref{smc}(b) suggests that the effect
is much reduced in the SMC as compared to, e.g. Figure \ref{gal}(b). One possible reason for this is that SMC Cepheids
have lower amplitudes than LMC or Galactic Cepheids \citep{pac00}.
Equation (1) shows that even if the range of
$T_{max}$ or $T_{min}$ is narrow, but if the amplitudes are not
large, then $T_{min}$ or $T_{max}$ will not be driven to such low or high values.

\subsection{Combining the PC and AC relations}

As discussed in the Introduction, equation (1) predicts that if the PC relation is flat at maximum light, then there will be an AC relation at minimum light, and vice versa. We see some evidence to support this idea from Figure \ref{gal}-\ref{smc}, and from Table \ref{tab1} \& \ref{tab2}. In summary:
\begin{description}
\item   Galactic short period Cepheids: PC relation steep at maximum $\rightarrow$ AC $\sim$flat at minimum.
\item   Galactic long period Cepheids: PC relation flat at maximum $\rightarrow$ AC relations $\sim$steep at minimum. PC relation steep at minimum $\rightarrow$ AC relation $\sim$flat at maximum.
\item   LMC short period Cepheids: PC relation $\sim$steep at maximum $\rightarrow$ AC relation flat at minimum.
\item   LMC long period Cepheids: PC relation flat at maximum $\rightarrow$ AC relations steep at minimum; PC relation steep at minimum $\rightarrow$ AC relation $\sim$flat at maximum.
\item   SMC short period Cepheids: PC relation steep at maximum $\rightarrow$ AC relation $\sim$flat at minimum.
\end{description}


\section {Conclusion and Discussion}

We have presented in Figures \ref{gal}-\ref{smc} new observational
characteristics of Cepheids. Following the work of \citet{tam02a}, we apply rigorous 
statistical tests to show that at mean light, there is a change in the LMC PC
relation between short ($\log(P)<1.0$) and long period Cepheids, as shown in Figure \ref{lmc}(a). In addition, the LMC data also exhibit a change in the PC relation at minimum light. However, we find no such change at mean and minimum light in the
Galactic and SMC Cepheids. We also find that the PC relations at minimum light generally show the smallest scatter among the three galaxies, as compare to the phases at maximum or mean light. Following the work of SKM we study amplitude-color diagrams at maximum, mean and 
minimum light and find very different behavior in the three galaxies for short and long period Cepheids. This difference not only occurs within a given galaxy between the short and long period Cepheids, but also occurs from galaxy to galaxy. For example, the behavior of
short and long period Cepheids in the LMC is very different (i.e. Figure \ref{lmc}), and the behavior of Galactic and LMC
Cepheids of short period is also very different. Thus we note that there is an effect with both period and metallicity. Further work with state of the art pulsation codes is under way to confront these observations with model calculations. In addition, the PC relations clearly show greater structure than a simple two line regression as used here. For example, the LMC data indicates another break at $\log (P)\sim 1.2$ (see, e.g., Figure \ref{residual}). This will be the subject of a future paper.

\subsection{Testing of Systematics}

In this subsection we discuss some of the systematic effects that may affect our results, as follows:
 
\begin{enumerate}
\item   Could reddening errors produce the results displayed in Figures \ref{gal}-\ref{smc}? Consider
the bottom panels of Figure \ref{gal}(b) and \ref{lmc}(b) showing AC relations at minimum light for the
Galaxy and LMC. In contrast to the Galactic counterpart,
the AC relation for short period LMC Cepheids is flat. It is difficult to
imagine a situation where the published reddening for these short period stars are in error  
to the extent of making this relation flat when it should have a slope like the Galactic
relation. Furthermore, we use the $E(B-V)$ values from the literature, which would not only affect our results but also other published results that are based on these values.

\item   Could outliers, perhaps stars that have been misclassified as Cepheids, produce
our significant results? We examine the OGLE LMC Cepheid data to investigate 
this question. The plot of mean $(V-I)_o$ color against $\log(P)$ in the middle panel of Figure \ref{lmc}(a) shows a number of
stars which seem to have very red colors. We exclude these stars
and repeat our analysis for PC relations in the LMC. We note that it is possible to
reduce the significance of our result at mean light by excluding stars that are
too red. However if we exclude stars that are too blue, the F test again becomes
significant. This conclusion can be seen in Table \ref{tab3}, where we apply various color cuts to the LMC data and then calculate the F test. Excluding the Cepheids with too red or blue color does not alter the results we have in Section 3.1.

\item   Could we reduce the significance of our results by ``judiciously'' removing certain stars?
Since we only consider stars with $\log (P) > 0.4$, if we extend this cutoff to 0.6 ($P$ = 3.98 days)
and repeat our analysis, we get very similar results. Also, excluding some of the longer period Cepheids,
those with periods greater than $\log (P) > 1.4$ increases the significance of our results. Hence the period cut we applied in this study would not significantly alter the results or conclusions we have.
\end{enumerate}


\begin{table}
\centering
\caption{F test significance results with color cuts for LMC Cepheids.}
\label{tab3}
\begin{tabular}{lcccccc} \\ \hline
$(V-I)_o$ & $F$ & $P(F)$ & $F$ & $P(F)$ & $F$ & $P(F)$ \\
Range     & Max   & Max      & Mean  & Mean     & Min   & Min    \\      
\hline
0 - 1.9 & 2.67 & 0.07 & 13.2 & 0.00 & 10.4 & 0.00 \\
0 - 1.1 & 2.95 & 0.05 & 14.0 & 0.00 & 10.7 & 0.00 \\
0 - 1.0 & 4.45 & 0.01 & 10.4 & 0.00 & 7.86 & 0.00 \\
0 - 0.9 & 3.34 & 0.04 & 7.69 & 0.00 & 6.38 & 0.00 \\
0.3 - 1.9 & 2.67 & 0.07 & 13.2 & 0.00 & 10.4 & 0.00 \\
0.4 - 1.9 & 2.48 & 0.08 & 14.8 & 0.00 & 12.2 & 0.00 \\
0.5 - 1.9 & 2.41 & 0.09 & 16.3 & 0.00 & 13.0 & 0.00 \\
0.5 - 0.9 & 3.16 & 0.04 & 10.4 & 0.00 & 8.76 & 0.00 \\
\hline
\end{tabular}
\end{table}

While excluding stars based on color or period cuts as mentioned above, in some cases, can reduce
the significance of our F test results at mean light (though never to a $P(F)$ value less than 0.1), such
experiments have very little effect at minimum light. If we accept
the assertion that mean light properties of Cepheids are an average of
Cepheid properties at all pulsation phases, then irrespective of mean light, Tables \ref{tab1}-\ref{tab2} and Figures
\ref{gal}-\ref{smc} clearly demonstrate a significant change in Cepheid observational properties between
short and long period Cepheids in the Galaxy, LMC and SMC.
Further if we accept these exclusions, they
would also have an effect on the PL relations in both V and I bands and hence on the currently
accepted extra-galactic distance scale.

\subsection{The period-luminosity relations in LMC}

Since the PL relation is affected by the PC relation, it is of interest to examine the PL relation in the OGLE LMC Cepheids, which show a break in the PC relation. Do the LMC data support a single PL relation or is the data consistent with a break at a period of 10 days as shown in \citet{tam02a}? We can use the F test to investigate the LMC PL relations in the V and I band as in Section 3. The results of the F test are shown in
Table \ref{tab4}, where column 1 displays the phase and
column 2 and 3 show the F statistic and its probability, under
the null hypothesis of a single line, for the V band PL relations. Similar results for I band PL relations are listed in column 4 and 5 in Table \ref{tab4}. From the table, both V and I band PL relations at mean and minimum
light are significant - that is the data are more consistent with a model where the slopes
of the V and I band PL relation are different for short and long period Cepheids.
However, the data are consistent with a single slope, in both V and I bands, for the
PL relation at maximum light, even though the corresponding PC relation shows some evidence of
a break at 10 days (see Table \ref{tab1}). 

The actual slopes of the V and I
band PL relation at maximum, mean and minimum light for short and long period
Cepheids are given in Table \ref{tab5}. The second column in this table gives the overall slope. The third and fourth
column give the short period and long period slopes, respectively. We clearly see
that the slope at maximum light is virtually identical for
short and long period Cepheids in both V and I bands. This is very different to the
situation at mean and minimum light. How can the PC relation at maximum light show evidence of a break
at $\log (P) = 1.0$ yet the PL relation in V and I be consistent with one PL relation? This is
certainly not the case for the PL relation at mean light in the LMC. This occurs because the
LMC PC relation at mean light becomes steeper for long period Cepheids whereas at maximum light,
the PC relation in the LMC becomes shallower for longer period Cepheids. Since the
PC relation affects the PL relation, this shallow dependence at maximum light for longer period
Cepheids suggests that the PL relation at that phase would not be affected too much. A detailed analysis of the PL relations at maximum, mean and minimum light will be presented in forthcoming paper.

\begin{table}
\centering
\caption{F test significance results for the LMC PL relation}
\label{tab4}
\begin{tabular}{lcccc} \\ \hline
phase & $F_{V\ band}$ & $P(F)_{V\ band}$ & $F_{I\ band}$ & $P(F)_{I\ band}$ \\ \hline
Max   & 0.32 & 0.73 & 0.04 & 0.96 \\
Mean  & 8.86 & 0.00 & 6.25 & 0.00 \\
Min   & 17.5 & 0.00 & 19.0 & 0.00 \\
\hline
\end{tabular}
\end{table}

\begin{table}
\centering
\caption{Slopes of the V and I band PL relation at maximum, mean and
minimum light for Cepheids in the LMC.}
\label{tab5}
\begin{tabular}{lcccccc} \\ \hline
Phase& V slope (all) & V slope (short) & V slope (long) \\ 
\hline
Max&$-2.74\pm0.05$&$-2.73\pm0.08$&$-2.97\pm0.26$\\
Mean&$-2.75\pm0.04$&$-2.95\pm0.07$&$-2.35\pm0.25$ \\
Min&$-2.59\pm0.04$&$-2.85\pm0.06$&$-1.84\pm0.26$& \\
\hline
Phase& I slope (all) & I slope (short) & I slope (long) \\ 
\hline
Max&$-2.95\pm0.03$&$-2.96\pm0.05$&$-2.94\pm0.19$ \\
Mean&$-2.99\pm0.03$&$-3.10\pm0.04$&$-2.94\pm0.18$ \\
Min&$-2.88\pm0.03$&$-3.06\pm0.04$&$-2.33\pm0.17$ \\
\hline
\end{tabular}
\end{table}

\subsection{Future work}

We new briefly summarize other implications from the present study. However, the detailed analysis of these topics are beyond the scope of this paper, and will be addressed in future papers.
 
\begin{enumerate}
\item   {\bf Metallicity dependence on PC relation}: From the period-mean density relation for a pulsator and the Stefan-Boltzmann law, it is easy to show that: $\log (P)=\alpha \log(L) + \beta \log (M) + \gamma \log (T_{eff}) + const$. The parameters ($\alpha,\beta,\gamma$) and the constant term can be determined from stellar pulsation calculations. For example:

\begin{description}
\item   $\log (P) = 0.84\log (L) - 0.68\log (M) - 3.48\log(T_{eff}) + const.$, from \citet{van73}.
\item   $\log (P) = 0.93\log (L) - 0.77\log (M) - 3.54\log(T_{eff}) + const.$, from \citet{sim97}.
\item   $\log (P) = 0.82\log (L) - 0.60\log (M) - 3.55\log(T_{eff}) + const.$, from \citet{bea01}.
\end{description}

In addition, the luminosity and mass should also obey an M-L relation from stellar evolution calculations. This M-L relation predicts that lower metallicity Cepheids will have higher luminosity (see, e.g., \citealt{bon00}) for given mass. Hence, at fixed period, lower metallicity Cepheids should be hotter than higher metallicity Cepheids. For example, \citet{lan86} showed that the long period SMC Cepheids are hotter than LMC Cepheids at given period by $\sim200 K$.  
 
Since the Magellanic Clouds (MC) have lower metallicity than the Galaxy, Cepheids in the MC should be bluer than Galactic Cepheids. \citet{tam03} have reported this finding in their paper. In this study, we also found that the short period MC Cepheids generally have bluer $(V-I)_o$ color than Galactic Cepheids, for a given period, at maximum, mean and minimum light. However, larger errors in the PC relations for long period Cepheids make such a statement more contentious for long period Cepheids. We intend to investigate this in detail in a future paper.  

\item   {\bf Estimation of color excess}: \citet{fer94} used the theory of SKM and the original suggestion of \citet{cod47} to establish a relation between $(B-V)$ color at maximum light, the $V$ band amplitude, period and the color excess $E(B-V)$ for Galactic Cepheids. Figure \ref{gal} shows that another approach would be using the properties of Cepheids at minimum light to estimate the color excess, since the PC relations at minimum light in generally show a smaller dispersion (see Section 3.1). Furthermore, the Galactic PC relation at maximum light indicates a break at 10 days (Table \ref{tab1}), but not in the PC relation at minimum light.
In future work, we plan to investigate the possibility of using this relation as a way to determine reddening.

\end{enumerate}


\section*{acknowledgements}
        We would like to thank Sergei Nikolaev and Gustav Tammann for useful discussion.


\end{document}